\documentclass[onecolumn,showkeys,showpacs,amsmath,amssymb]{revtex4}
\input{epsf}

\usepackage{graphicx}
\usepackage{array}
\usepackage{bigstrut}
\usepackage{longtable}
\usepackage{rotating,booktabs}
\usepackage{booktabs,threeparttable}
\usepackage{bm}
\usepackage{lipsum}
\usepackage{multirow}
\usepackage{color}
\usepackage[colorlinks,linkcolor=blue,anchorcolor=blue,citecolor=blue,urlcolor=blue]{hyperref}

\begin{document}

\title{Relativistic hyperpolarizabilities for atomic H, Li, and Be$^+$ systems}

\author{Shan-Shan Lu,$^{1,2}$ Hong-Yuan Zheng,$^{1,2}$ Zong-Chao Yan,$^{3,1}$ James F. Babb,$^{4}$ and Li-Yan Tang$^{1,*}$~\footnotetext{* Email Address: lytang@apm.ac.cn}}

\affiliation {$^1$ State Key Laboratory of Magnetic Resonance and Atomic and Molecular Physics, Wuhan Institute of Physics and Mathematics, Innovation Academy for Precision Measurement Science and Technology, Chinese Academy of Sciences, Wuhan 430071, People's Republic of China}

\affiliation {$^2$ University of Chinese Academy of Sciences, Beijing 100049, People's Republic of China}

\affiliation {$^3$ Department of Physics, University of New Brunswick, Fredericton, New Brunswick, Canada E3B 5A3}

\affiliation {$^4$ ITAMP, Center for Astrophysics \textbar\ Harvard \& Smithsonian, 60 Garden St., Cambridge, MA 02138, USA}

\date{\today}

\begin{abstract}
For atoms in external electric fields, the hyperpolarizabilities are the coefficients describing the nonlinear interactions contributing to the induced energies at the fourth power of the applied electric fields. Accurate evaluations of these coefficients for various systems are crucial for improving precision in advanced atom-based optical lattice clocks and for estimating field-induced effects in atoms for quantum information applications. However, there is a notable scarcity of research on atomic hyperpolarizabilities, especially for the relativistic realm. Our work addresses this gap by establishing a novel set of alternative formulas for the hyperpolarizability based on fourth-order perturbation theory. These formulas offer a more reasonable regrouping of scalar and tensor components compared to previous formulas, thereby enhancing their correctness and applicability. To validate our formulas, we perform the calculations for the ground and low-lying excited pure states of few-electron atoms H, Li, and Be$^+$. The highly accurate results obtained for the H atom could serve as benchmarks for further development of other theoretical methods.
\end{abstract}



\maketitle

\section{Introduction}
The response of an atom to an applied electric field is described at leading order by the polarizability (linear response to the field) and at higher order by the hyperpolarizability (nonlinear response to the field). The literature on precision calculations of polarizabilities for atoms~\cite{mitory10,tang12b,zhang16,zhang16b,tang10a} is extensive compared to that for hyperpolarizabilities, but the methods used to calculate polarizabilities are not necessarily directly applicable to hyperpolarizabilities. Here, our intent is to present a theoretical formalism and some highly-accurate benchmark values for hyperpolarizabilities of few-electron atoms.

Hyperpolarizabilities~\cite{bishop95a,lupinetti06b,shelton94} are the coefficients describing the response induced in atoms at the order of the fourth power of applied electric fields (and describe induced dipole moments at the third power of applied electric fields). As perturbative corrections to the energies of atoms in electric fields, the relative values of the hyperpolarizabilities compared to the values of the polarizabilities (which appear at the second power of applied electric fields) are fundamental to estimating the impact of field intensities on nonlinear phenomena~\cite{manakov86b,manakov86c}. Knowledge of the values of hyperpolarizabilities plays a crucial role in arrangements to suppress nonlinear effects for precision laser spectroscopy measurements~\cite{henson22,yang16,sheng18,guo20}, especially in the pursuit and development of ever higher-precision optical clocks~\cite{nicholson15,brown17,ushijima18,kim23}. Moreover, applications of highly excited Rydberg states in neutral atom quantum computing are rapidly developing and it is reasonable to expect that improvements in quantification of hyperpolarizabilities will be needed as control of these atoms in electric fields is sought  [see, for example, Eq.~(11) of Ref.~\cite{bohorquez2023}].

In this study, we present a set of general formulas for the atomic hyperpolarizabilities, which improves our previously published theory~\cite{tang09a} by offering a more reasonable regrouping of the scalar and tensor components, thus preserving the common understanding of the relativistic correction being small. The utility of the new formulas is demonstrated by conducting high-accuracy calculations on H, Li, and Be$^+$ using both nonrelativistic and relativistic approaches. For the H atom in an excited state, we exclude all degenerate states when calculating the hyperpolarizabilities~\cite{jhanwar82}. These so-called pure state hyperpolarizabilities can serve as benchmarks for testing computational methods. For the Li and Be$^+$ systems, we leverage the nonrelativistic numerical values obtained using our previous theory~\cite{tang09a,tang09b}, demonstrating that the correctness and applicability of present formulas.

Atomic units (a.u.), where $e=1$, $a_0=1$,  and $\hbar=1$, are used throughout; the atomic unit of the (second) hyperpolarizability is  $e^4a_0^4/E_{\mathrm{h}}^3$ or about $6.235\,379\,9905(38) \times 10^{-65}\, \mathrm{C}^4\, \mathrm{m}^4\, \mathrm{J}^{-3}$, where $e$ is the electron charge, $a_0$ is the Bohr radius, and $E_{\mathrm{h}}=e^2/a_0$ is the Hartree.

\section{An alternative formula}\label{formalism}

When an atom is placed in an uniform external electric field, its energy levels change due to the influence of the field. Specifically, the energy at second-order in the electric field depends on the atomic dipole polarizability. The fourth-order shift depends on the atomic hyperpolarizability $\gamma$ and can be expressed in the form:
\begin{eqnarray}
	\Delta E_4=-\frac{\varepsilon^4}{24}\gamma =-\frac{\varepsilon^4}{24}
	\bigg\{\gamma_0+g_2(J,M_J)\gamma_2+\big[g_4(J,M_J)\gamma_4^{(1)}+g_2^2(J,M_J)\gamma_4^{(2)}\big]\bigg\}\,,\label{e1}
\end{eqnarray}
where $\boldsymbol{\varepsilon}$ is the electric field strength, the coefficients $g_2(J,M_J)$ and $g_4(J,M_J)$ depend on the magnetic quantum number $M_J$,
\begin{eqnarray}
	g_2(J,M_J)=
	\begin{cases}
		0,                            & J \leq \frac{1}{2}\,, \\
		\frac{3M_J^2-J(J+1)}{J(2J-1)},& {\rm otherwise}\,,      \label{e2}
	\end{cases}
\end{eqnarray}
\begin{eqnarray}
	g_4(J,M_J)=
	\begin{cases}
		0,                            & J \leq \frac{3}{2}\,, \\
		\frac{3(5M_J^2-J^2-2J)(5M_J^2+1-J^2)-10M_J^2(4M_J^2-1)}{J(2J-1)(2J-2)(2J-3)},& {\rm otherwise}\,.    \label{e3}
	\end{cases}
\end{eqnarray}
Also in Eq.~(\ref{e1}), $\gamma_0$ corresponds to the scalar component of the  hyperpolarizability, whereas $\gamma_2$, $\gamma_4^{(1)}$, and $\gamma_4^{(2)}$ correspond to the tensor components, which have the following forms:
\begin{eqnarray}
	\gamma_0&=&(-1)^{2J}\frac{24}{\sqrt{2J+1}}  \sum\limits_{J_aJ_bJ_c}\mathcal{G}_0^{(1)}(J,J_a,J_b,J_c)\mathcal{T}_1(J_a,J_b,J_c)-\frac{24}{2J+1}\sum\limits_{J_aJ_c}\mathcal{G}_{00}^{(2)}(J,J_a,J_c) \mathcal{T}_2(J_a,J_c)\,,  \label{e4}
\end{eqnarray}
\begin{eqnarray}
	\gamma_2&=&(-1)^{2J} 24\sqrt{\frac{J(2J-1)}{(2J+3)(J+1)(2J+1)}} \sum\limits_{J_aJ_bJ_c}\mathcal{G}_2^{(1)}(J,J_a,J_b,J_c)\mathcal{T}_1(J_a,J_b,J_c)\nonumber \\
	&-&\frac{24}{2J+1} \sqrt{\frac{J(2J-1)}{(2J+3)(J+1)}}     \sum\limits_{J_aJ_c} [\mathcal{G}_{02}^{(2)}(J,J_a,J_c)+\mathcal{G}_{20}^{(2)}(J,J_a,J_c)]  \mathcal{T}_2(J_a,J_c)\,, \label{e5}
\end{eqnarray}
\begin{eqnarray}
	\gamma_4^{(1)}&=&(-1)^{2J} 24\sqrt{\frac{J(2J-1)(J-1)(2J-3)}{(2J+5)(J+2)(2J+3)(J+1)(2J+1)}}  \sum\limits_{J_aJ_bJ_c}\mathcal{G}_4^{(1)}(J,J_a,J_b,J_c)\mathcal{T}_1(J_a,J_b,J_c)\,,   \label{e6}
\end{eqnarray}
and
\begin{eqnarray}
	\gamma_4^{(2)}&=&-\frac{24J(2J-1)}{(2J+3)(2J+1)(J+1)} \sum\limits_{J_aJ_c}\mathcal{G}_{22}^{(2)}(J,J_a,J_c)  \mathcal{T}_2(J_a,J_c)\,,  \label{e7}
\end{eqnarray}
where $\mathcal{T}_1(J_a,J_b,J_c)$ and $\mathcal{T}_2(J_a,J_c)$ are defined in  Appendix~\ref{stark_a} by Eqs.~(\ref{a9}) and (\ref{a10}), $\mathcal{G}_\lambda^{(1)}(J,J_a,J_b,J_c)$ with $\lambda=0,2,4$ and $\mathcal{G}_{k_1k_2}^{(2)}(J,J_a,J_c)$ are defined in Appendix~\ref{stark_a} by Eqs.~(\ref{a18}) and (\ref{a19}).
The detailed derivation of Eq.~(\ref{e1}) is provided in Appendix~\ref{stark_a}.
This formula Eq.~(\ref{e1}) features a distinct partitioning, in terms of both scalar and tensor components, compared to the commonly used one in Refs.~\cite{manakov86c, tang09a}, while still yielding the same total hyperpolarizability $\gamma$.

\section{Computational Methods}
\subsection{Single-electron Schr\"{o}dinger and Dirac equations}

When solving the hydrogen, the nuclear mass is set to be infinite, and the point nucleus model is adopted.
For the nonrelativistic case, the Schr\"{o}dinger equation for hydrogen is given by
\begin{eqnarray}
	\left[-\textstyle{\frac{1}{2}}\nabla^2+V(r)\right]\psi_S(\bm{r})=E_S\psi_S(\bm{r)}\,, \label{e35}
\end{eqnarray}
where $V(r)=-Z/r$ and $Z$ is the nuclear charge. Due to its rotational symmetry, there exist solutions to Eq.~(\ref{e35}) of the separable form
\begin{eqnarray}
	\psi_S(\bm{r})=U_{nl}(r)Y_{lm}(\theta,\phi)\,, \label{e36}
\end{eqnarray}
where $Y_{lm}(\theta,\phi)$ is a spherical harmonic function and the radial function $U_{nl}(r)$ satisfies the differential equation
\begin{eqnarray}
	\left[-\frac{1}{2}\frac{d^2}{dr^2}-\frac{1}{r}\frac{d}{dr}+\frac{l(l+1)}{2r^2}+V(r)\right]U_{nl}(r)=E_SU_{nl}(r)\,. \label{e37}
\end{eqnarray}

For the relativistic case, the Dirac equation for hydrogen is given by
\begin{eqnarray}
	\big[c\bm{\alpha}\cdot\bm{p}+\beta c^2+V(r)\big]\psi_D(\bm{r})=E_D\psi_D(\bm{r)}\,, \label{e38}
\end{eqnarray}
where $c=137.035\,999\,084$ is the speed of light~\cite{tiesinga21}, $\bm{p}$ is the momentum operator, and $\bm{\alpha}$ and $\beta$ are the usual $4\times4$ Dirac matrices~\cite{kaneko77a}.
The solutions to Eq.~(\ref{e38}) can be expressed in the following separable form:
\begin{eqnarray}
	\psi_D(\bm{r})=\frac{1}{r}\begin{pmatrix} iP_{n\kappa}(r)\Omega_{\kappa m}(\hat{\bf r})\\ Q_{nk}(r)\Omega_{-\kappa m}(\hat{\bf r}) \end{pmatrix}\,,  \label{e40}
\end{eqnarray}
where $P_{n\kappa}(r)$ and $Q_{n\kappa}(r)$ are the large and small components of the radial function, and $\Omega_{\kappa m}(\hat{\bf r})$ and $\Omega_{-\kappa m}(\hat{\bf r})$ are the angular components.
By substituting Eq.~(\ref{e40}) into Eq.~(\ref{e38}), we obtain the following coupled first-order differential equations that describe the behavior of $P_{n\kappa}(r)$ and $Q_{n\kappa}(r)$:
\begin{eqnarray}
	\begin{pmatrix}V(r)&c\left(\frac{d}{dr}-\frac{\kappa}{r}\right)\\ -c\left(\frac{d}{dr}+\frac{\kappa}{r}\right)&-2c^2+V(r)\end{pmatrix} \begin{pmatrix} P_{n\kappa}(r)\\ Q_{n\kappa}(r) \end{pmatrix} &=&E\begin{pmatrix} P_{n\kappa}(r)\\ Q_{n\kappa}(r) \end{pmatrix}\,,
	\label{e41}
\end{eqnarray}
where $E=E_D-c^2$.

While it is possible to solve the Schr\"{o}dinger equation Eq.~(\ref{e37}) and the Dirac equation Eq.~(\ref{e41}) analytically, in this study, we choose to solve them numerically using $B$-spline functions within a specific cavity~\cite{johnson88a,fischer93,bachau01a,tang12b}. This approach enables us to validate our numerical method, which will be subsequently used to diagonalize the Hamiltonian.
For example, by using the complete $B$-spline basis, we can express $U_{nl}(r)$ as a discrete sum of $N$-dimensional, $k$-order $B$-spline basis functions:
\begin{eqnarray}
	U_{nl}(r)=\sum\limits_{i=1}^Nc_iB_i^k(r)\,. \label{e42}
\end{eqnarray}
The function $B_i^k(r)$ is non-zero only within the knot intervals $t_i \leq r \leq t_{i+k}$, where ${t_i}$ represents the knot sequence. In the nonrelativistic case, the radial wave functions must satisfy the boundary conditions $U_{nl}(0)=U_{nl}(R)=0$, where $R$ is the radius of the cavity. In the relativistic case, the Notre Dame (ND) boundary conditions~\cite{johnson88a,chodos74a} of $P_{n\kappa}(0)=0$ and $P_{n\kappa}(R)=Q_{n\kappa}(R)$ are imposed to address the issues related to the ``Klein Paradox".
In order to obtain accurate bound-state wave functions, it is more appropriate to use an exponential knot distribution for $B_i^k(r)$ given by
\begin{eqnarray}
	t_{i+k-1} = R\times\frac{\mathrm{exp}\left[\eta\left(\frac{i-1}{n-1}\right)\right]-1}{\mathrm{exp}(\eta)-1}\,,\ \ i=1,2,\ldots, N-k+2\,.  \label{e43}
\end{eqnarray}
Here, $\eta=a \times R$ is an adjustable exponential knot parameter. In the present work, the radii of the confining cavity are fixed at $R=400$~a.u. and $R=600$~a.u. for nonrelativistic and relativistic calculations, respectively. The value of $a$ is optimized by reproducing the exact energies~\cite{bethe77a} of low-lying states up to 20 significant digits.

\subsection{DFCP method}\label{DFCP}
Since the Li atom and Be$^+$ ion can be regarded as a frozen core combined with one valence electron, these systems can be solved using the Dirac-Fock plus core polarization (DFCP) method~\cite{johnson88a, bachau01a}. As described in previous studies~\cite{tang13b, jiang16, wu19}, the initial step in the DFCP method is to perform a Dirac-Fock calculation of the frozen core. For monovalent electron systems, the effective Hamiltonian can be written as:
\begin{equation}
	h_{\rm DFCP}(\bm{r})=c{\bm{\alpha}}\cdot{\bm{p}}+(\beta-1)c^{2}+V(r)+V_{\rm dir}(r)+V_{\rm exc}(r)+V_{\rm 1}(r)\,. \label{e48}
\end{equation}
Here, $V_{\rm dir}(r)$ and $V_{\rm exc}(r)$ are the direct and exchange potentials between the core electrons and the valence electron, respectively, and $V_{\rm 1}(r)$ is the semiempirical polarization potential,
\begin{eqnarray}
	V_{1}(r)=-\frac{\alpha_{\rm core}}{2r^4}\bigg[1-{\rm exp}\bigg(-\frac{r^6}{\rho_\kappa^6}\bigg)\bigg]\,, \label{e49}
\end{eqnarray}
where $\alpha_{\rm core}$ is the core dipole polarizability, $\rho_\kappa$ is the cutoff parameter, and $\kappa$ is the angular quantum number. In the present work, we used $\alpha_{\rm core}=0.1894$~a.u. for Li$^+$ and $\alpha_{\rm core}= 0.05182$~a.u. for Be$^{2+}$~\cite{johnson83}. The cutoff parameter $\rho_\kappa$ is fine-tuned to accurately replicate the experimental energy of the lowest state associated with each value of $\kappa$.

\section{Results and Discussions}
\subsection{Hydrogen}

Table~\ref{t1} presents a convergence analysis of the nonrelativistic and relativistic hyperpolarizabilities, $\gamma_0$, for the ground state of the hydrogen atom. As the number of $B$-spline basis functions $N$ varies from 100 to 400, the nonrelativistic value gradually approaches the exact value of 10665/8=1333.125~a.u.~\cite{bishop92}. Additionally, the relativistic result has been accurately determined to 21 significant figures. The extrapolated value can be obtained by assuming that the ratio of two successive hyperpolarizability differences remains constant as $N$ tends to infinity. Comparing the nonrelativistic and relativistic values, we find that the relativistic correction to the ground-state hyperpolarizability is $-0.135\,342\,240\,716\,897\,63$~a.u.

The convergence of calculations for the nonrelativistic $2p$ and $3d$ states, as the size of the $B$-spline basis set $N$ increases, is presented in Tables~\ref{ta1} and \ref{ta2} of Appendix~\ref{table_a}. It can be observed that the scalar and tensor components have converged to at least 21 significant digits. Specifically, for the $2p\, (|M_L|=1)$ state, the total hyperpolarizability $\gamma$ is determined as $5\,326\,848$~a.u. by utilizing Eqs.~(\ref{e1})-(\ref{e3}), which aligns perfectly with the result reported in Ref.~\cite{cohen06a}. It is worth reiterating that in our calculations, we have excluded all degenerate states by considering only the pure state hyperpolarizabilities~\cite{jhanwar82}, as shown in Eqs. (\ref{a9}) and (\ref{a10}) of Appendix~\ref{stark_a}.

In the relativistic case, the hyperpolarizabilities of $2p_{1/2}$ and $2p_{3/2}$ states have been calculated up to 21 significant digits, as shown in Table~\ref{ta3} of Appendix~\ref{table_a}. These values, when compared to the nonrelativistic results, have decreased due to the relativistic corrections. Specifically, the relativistic correction for the scalar component $\gamma_0$ of the $2p_{1/2}$ state is $-1079.819\,152\,781\,342\,11$~a.u., while for the $2p_{3/2}$ state it is $-58.946\,729\,464\,319\,25$~a.u. Additionally, for the tensor components $\gamma_2$ and $\gamma_4^{(2)}$ of the $2p_{3/2}$ state, these corrections are $-93.818\,725\,858\,667\,39$~a.u. and $-1.166\,186\,367\,213\,346\,6$~a.u., respectively.
For the $3d_{3/2}$ and $3d_{5/2}$ states, as presented in Tables~\ref{ta4} and~\ref{ta5} of Appendix~\ref{table_a}, the hyperpolarizabilities have been determined with an accuracy of more than 19 significant digits. The differences relative to the nonrelativistic $\gamma_0$ are $-42\,365.011\,332\,071\,4$~a.u. for the $3d_{3/2}$ state and $-10\,610.289\,079\,484\,2$~a.u. for the $3d_{5/2}$ state. Notably, the tensor components $\gamma_2$, $\gamma_4^{(1)}$, and $\gamma_4^{(2)}$ of the $3d_{5/2}$ state closely approximate the nonrelativistic results. All the final values of $\gamma_0$, $\gamma_2$, $\gamma_4^{(1)}$, and $\gamma_4^{(2)}$ are summarized in Table~\ref{t2}.

\begin{table*}[!htbp]
\caption{\label{t1} Convergence of the nonrelativistic and relativistic hyperpolarizability $\gamma_0$ for the ground state of the H atom, as the size of basis set $N$ increases progressively, in atomic units. It should be noted that $N$ represents the number of intermediate states with different quantum numbers, and since these numbers are all the same, we use only $N$ to uniformly identify them.}
\begin{ruledtabular}
\begin{tabular}{llll}
\multicolumn{2}{c}{Nonrelativistic}&\multicolumn{2}{c}{Relativistic}\\
\cline{1-2}\cline{3-4}
\multicolumn{1}{l}{$N$}&\multicolumn{1}{c}{$\gamma_0$}&\multicolumn{1}{l}{$N$}&\multicolumn{1}{c}{$\gamma_0$}\\  \hline
100&    1 333.125 000 000 004           &  100&     1 332.989 657 756 \\
150&    1 333.125 000 000 000 023       &  200&     1 332.989 657 759 283 081 \\
200&    1 333.125 000 000 000 000 65    &  300&     1 332.989 657 759 283 102 377\\
250&    1 333.125 000 000 000 000 041   &  400&     1 332.989 657 759 283 102 385\\
300&    1 333.125 000 000 000 000 004 4 &  500&     1 332.989 657 759 283 102 376\\
350&    1 333.125 000 000 000 000 000 67&  600&     1 332.989 657 759 283 102 372\\
400&    1 333.125 000 000 000 000 000 13&  Extrap.& 1 332.989 657 759 283 102 370\\
Extrap.&1 333.125 000 000 000 000 000 05&         &  \\
Ref.~\cite{bishop92}&   1 333.125       &         &  \\
\end{tabular}
\end{ruledtabular}
\end{table*}
\begingroup
\squeezetable
\begin{table*}[!htbp]
\caption{\label{t2} Summary of nonrelativistic and relativistic pure state hyperpolarizabilities for the low-lying states of the H atom, in atomic units. All the tabulated digits are insensitive to further enlargement of the $B$-spline basis functions. The numbers in the square brackets denote powers of ten.}
\begin{ruledtabular}
\begin{tabular}{lllll}
\multicolumn{1}{l}{State}&\multicolumn{1}{c}{$\gamma_0$}&\multicolumn{1}{c}{$\gamma_2$}&\multicolumn{1}{c}{$\gamma_4^{(2)}$}&\multicolumn{1}{c}{$\gamma_4^{(1)}$}\\  \hline
\multicolumn{5}{c}{Nonrelativistic}\\
$1s$&   1333.125            &                          &                    &                       \\
$2p$&   8.130 560[6]        &   $-$2.769 472[6]        &   $-$3.424[4]      &                       \\
$3d$&   1.913 524 179 3[9]  &   $-$1.027 110 959 357 142 857 142[9]&   $-$6.944 162 4[7]&    2.939 248 330 714 285 714 28[7]\\  \hline
\multicolumn{5}{c}{Relativistic}\\
$1s_{1/2}$&  1 332.989 657 759 283 102 37   &                                    &                                    &   \\
$2p_{1/2}$&  8.129 480 180 847 218 657 89[6]&                                    &                                    &   \\
$2p_{3/2}$&  8.130 501 053 270 535 680 75[6]&  $-$2.769 565 818 725 858 667 39[6]&  $-$3.424 116 618 636 721 334 66[4]&   \\
$3d_{3/2}$&  1.913 481 814 288 667 928 6[9] &  $-$7.189 813 879 847 891 582 0[8] &  $-$3.402 556 953 695 351 845 4[7] &   \\
$3d_{5/2}$&  1.913 513 569 010 920 515 8[9] &  $-$1.027 113 325 161 421 946 0[9] &  $-$6.944 098 825 526 114 424 3[7] &  2.939 326 812 562 169 711 9[7]\\
\end{tabular}
\end{ruledtabular}
\end{table*} 
\endgroup
\subsection{Li atom and Be$^+$ ion}

We have recalculated the nonrelativistic hyperpolarizabilities of the Li atom and Be$^+$ ion in their low-lying states, in the framework of the Hylleraas variational method using the formulas presented in Sec.~2. More information on the computational methodology and convergence studies can be found in our previous publications~\cite{tang09a,tang09b,tang10a}. For the relativistic hyperpolarizabilities, we employed the DFCP method. Specifically, we performed single-electron calculations for the relativistic hyperpolarizabilities of Li and Be$^+$ using a frozen-core Hamiltonian and a semiempirical polarization potential as detailed in Sec.~3.2. To ensure accuracy, we substituted our energies with the corresponding NIST data~\cite{nist22}, resulting in errors primarily originating from the theoretical reduced matrix elements. By comparing these matrix elements with the values in Ref.~\cite{UDportal22}, we observed relative differences within 0.3\% and 0.2\% for Li and Be$^+$, respectively. To conservatively estimate the uncertainties in the relativistic hyperpolarizabilities, we introduced fluctuations of 0.3\% and 0.2\%, respectively, into all the reduced matrix elements. The present nonrelativistic (Hylleraas) and relativistic (DFCP) results are listed in Tables~\ref{t3} and~\ref{t4}.

From the comparison of Hylleraas and DFCP results presented in Tables~\ref{t3} and~\ref{t4}, it is clear that they generally agree with each other except for the ground state of Li atom. There exists a significant discrepancy between the DFCP result of 5750~a.u. and the Hylleraas result of 3060(40)~a.u., which can be attributed to a numerical cancellation of DFCP result.
In the case of the hyperpolarizability of the $2\,s_{1/2}$ state, it is determined as the sum of nine distinct types of intermediate states. These intermediate states can be further categorized based on their orbital angular quantum numbers, denoted as $(m, n, k)=(mp_j, ns_{1/2}, kp_j)$ and $(mp_j, nd_{j'}, kp_j)$, where $j=1/2, 3/2$, $j'=3/2, 5/2$, and $\left| j-j' \right| \leq 1$. The contributions from $(mp_j, ns_{1/2}, kp_j)$ and $(mp_j, nd_{j'}, kp_j)$ are $-1\,469\,773$~a.u. and $1\,475\,523$~a.u., respectively. As a result of this cancellation, there is a loss of three significant figures when combining the two contributions to obtain the total $\gamma_0$. Consequently, we can only provide the central value of 5750~a.u., which has the same order of magnitude as the nonrelativistic result. However, this situation does not arise for the ground state of Be$^+$. The contributions from $(mp_j, ns_{1/2}, kp_j)$ and $(mp_j, nd_{j'}, kp_j)$ amount to $-20263$~a.u. and $8755$~a.u., respectively, and their sum equals $-11508(92)$~a.u., which is in good agreement with the all-order result of $-11496(6)$~a.u. as reported in Ref.~\cite{safronova13e}.

\begin{table*}[!htbp]
\caption{\label{t3} Nonrelativistic and relativistic hyperpolarizabilities of the low-lying states of the Li atom, in atomic units. The numbers in parentheses denote computational uncertainty, while the numbers in square brackets represent the power of 10.}
\begin{ruledtabular}
\begin{tabular}{lllll}
\multicolumn{1}{l}{State}&\multicolumn{1}{c}{$\gamma_0$}&\multicolumn{1}{c}{$\gamma_2$}&\multicolumn{1}{c}{$\gamma_4^{(2)}$}&\multicolumn{1}{c}{$\gamma_4^{(1)}$}\\  \hline
\multicolumn{5}{c}{Nonrelativistic (Hylleraas)}\\
$2s$                &       3060(40)  &                    &                 &                  \\
$2p$                &   9.9854(4)[6]  &	 $-$6.2074(4)[6]   &	 1.5539(2)[4]&                  \\
$3d$                &   9.38618(1)[11]&	 $-$1.93402(1)[12] &   1.01218(1)[12]&	$-$1.5938(6)[10]\\  \hline
\multicolumn{5}{c}{Relativistic (DFCP)}\\
$2s_{1/2}$&           5750&                   &               &                   \\
$2p_{1/2}$&    9.82(12)[6]&                   &               &                   \\
$2p_{3/2}$&    9.82(12)[6]&    $-$6.126(74)[6]&   1.439(17)[4]&                   \\
$3d_{3/2}$&   9.37(12)[11]&   $-$1.351(17)[12]&  4.948(60)[11]&                   \\
$3d_{5/2}$&   9.37(12)[11]&   $-$1.930(23)[12]&  1.009(13)[12]&   $-$1.582(19)[10]\\
\end{tabular}
\end{ruledtabular}
\end{table*}
\begin{table*}[!htbp]
\caption{\label{t4} Nonrelativistic and relativistic hyperpolarizabilities of the low-lying states of the Be$^+$ ion, in atomic units. The numbers in parentheses denote computational uncertainty, while the numbers in square brackets represent the power of 10.}
\begin{ruledtabular}
\begin{tabular}{lllll}
\multicolumn{1}{l}{State}&\multicolumn{1}{c}{$\gamma_0$}&\multicolumn{1}{c}{$\gamma_2$}&\multicolumn{1}{c}{$\gamma_4^{(2)}$}&\multicolumn{1}{c}{$\gamma_4^{(1)}$}\\  \hline
\multicolumn{5}{c}{Nonrelativistic (Hylleraas)}\\
$2s$                &  $-$11521.30(3)&                  &               &                 \\
$2p$                &   6.64785(2)[3]&  $-$5.24075(4)[3]&  2.13286(2)[3]&                 \\
$3d$                &   5.9851(2)[8] &  $-$1.3038(4)[9] &  7.6397(2)[8] &  $-$5.7795(4)[7]\\ \hline
\multicolumn{5}{c}{Relativistic (DFCP)}\\
$2s_{1/2}$&     $-$11508(92)&                 &               &                 \\
Ref.~\cite{safronova13e}&  $-$11496(6)&       &               &                 \\
$2p_{1/2}$&     6.647(54)[3]&                 &               &                 \\
$2p_{3/2}$&     6.650(53)[3]&  $-$5.237(42)[3]&   2.134(17)[3]&                 \\
$3d_{3/2}$&     5.952(48)[8]&  $-$9.081(73)[8]&   3.730(30)[8]&                 \\
$3d_{5/2}$&     5.966(48)[8]&  $-$1.301(11)[9]&   7.638(61)[8]&  $-$5.835(47)[7]\\
\end{tabular}
\end{ruledtabular}
\end{table*}
\subsection{Comparison of two sets of formulas}

\begin{figure}
	\includegraphics[width=0.9\textwidth]{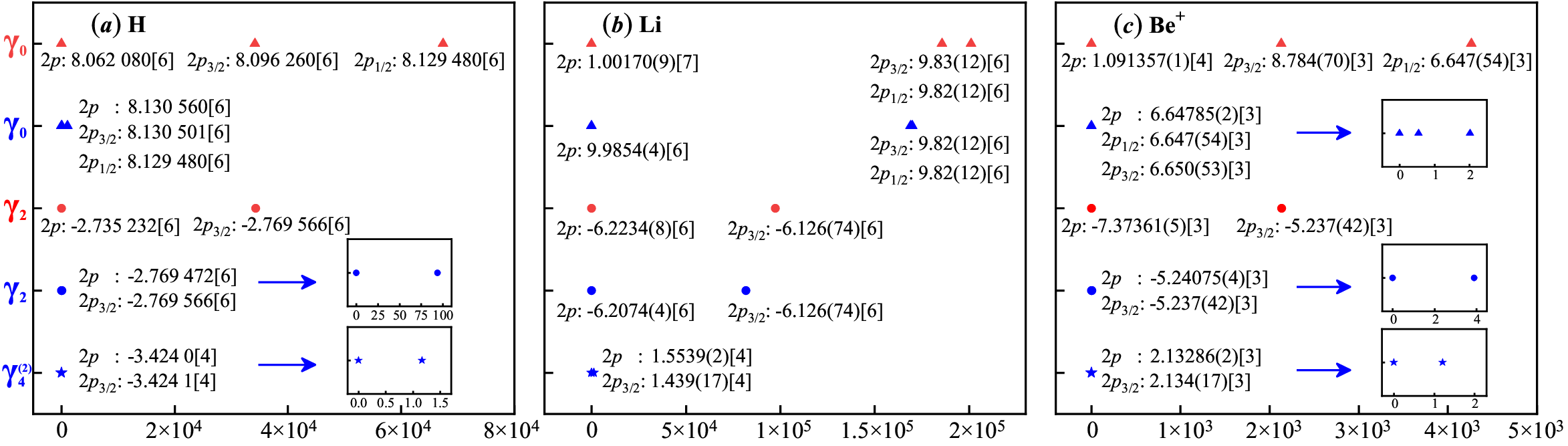}
	\caption{\label{f1}(Color online) Comparison of the present and previous~\cite{tang09a} relativistic corrections on the scalar and tensor hyperpolarizabilities for the $2p$ state of H, Li and Be$^+$, where the values calculated using the previous formula from Ref.~\cite{tang09a} are shown in red, while the values calculated using the present formula Eq.~(\ref{e1}) are shown in blue. Also $\gamma_0$, $\gamma_2$, and $\gamma_4^{(2)}$ are denoted by filled triangles, circles, and stars, respectively. All the data are referenced to the nonrelativistic results on the far left, and the horizontal axis represents the absolute value of relativistic corrections. The inserts provide a zoomed-in view of the overlapping points within the main image.}
\end{figure}

The nonrelativistic and relativistic scalar and tensor components of hyperpolarizability, as presented in Tables~\ref{t2}, \ref{t3} and \ref{t4}, exhibit minimal differences when calculated using the current formula in Eq.~(\ref{e1}), as generally accepted that the relativistic effects are small. In contrast, for the previous formula described in Refs.~\cite{manakov86c,tang09a}, this does not hold true, as demonstrated in Figure~\ref{f1} for the $2p$ state of Li and Be$^+$, although the total hyperpolarizability $\gamma$ remains the same.

\begingroup
\squeezetable
\begin{table*}[!htbp]
\caption{\label{t5} Comparison of hyperpolarizabilities for the nonrelativistic $2p$ and relativistic $2p_{3/2}$ states of the Be$^+$ ion, performed using the present formula Eq.~(\ref{e1}) and the previous formula Eq. (27) of Ref.~\cite{tang09a}. The numbers in parentheses indicate the computational uncertainty. In atomic units.}
\begin{ruledtabular}
\begin{tabular}{lcccccccccc}
\multirow{2}*{State}&\multirow{2}*{}&\multirow{2}*{$\gamma_0$}&\multirow{2}*{$\gamma_2$}&\multirow{2}*{$\gamma_4^{(2)}$}&\multicolumn{2}{c}{$g_2$}&\multicolumn{2}{c}{$g_2^2$}&\multirow{2}*{$\gamma(M_L=0)$}&\multirow{2}*{$\gamma(|M_L|=1)$}\\ \cline{6-7}  \cline{8-9}
\multicolumn{1}{c}{}&\multicolumn{1}{c}{}&\multicolumn{1}{c}{}&\multicolumn{1}{c}{}&\multicolumn{1}{c}{}&\multicolumn{1}{c}{$M_L=0$}&\multicolumn{1}{c}{$|M_L|=1$}&\multicolumn{1}{c}{$M_L=0$}&\multicolumn{1}{c}{$|M_L|=1$}&\multicolumn{1}{c}{}&\multicolumn{1}{c}{}\\ \hline
$2p$&  Present                &  6647.85(2) &  $-$5240.75(4)&  2132.86(2)& $-$2&  1&  4&  1&  25660.79(5)&  3539.96(5)\\
	&  Previous~\cite{tang09a}&  10913.57(1)&  $-$7373.61(5)&            & $-$2&  1&   &   &  25660.79(5)&  3539.96(5)\\
			
\multicolumn{1}{c}{}&\multicolumn{1}{c}{}&\multicolumn{1}{c}{}&\multicolumn{1}{c}{}&\multicolumn{1}{c}{}&\multicolumn{1}{c}{$|M_J|=1/2$}&\multicolumn{1}{c}{$|M_J|=3/2$}&\multicolumn{1}{c}{$|M_J|=1/2$}&\multicolumn{1}{c}{$|M_J|=3/2$}&\multicolumn{1}{c}{$\gamma(|M_J|=1/2)$}&\multicolumn{1}{c}{$\gamma(|M_J|=3/2)$}\\ \hline
$2p_{3/2}$&  Present                &  6650(53)&  $-$5237(42)&  2134(17)& $-$1&  1&  1&  1&  14021(70)&  3547(70)\\
	      &  Previous~\cite{tang09a}&  8784(70)&  $-$5237(42)&          & $-$1&  1&   &   &  14021(82)&  3547(82)\\
\end{tabular}
\end{ruledtabular}
\end{table*}
\endgroup

The numerical exemplification pertinent to the aforementioned point can be referenced by examining the values of Be$^+$ in the $2p$ state, as presented in Table~\ref{t5}. According to the previous formula~\cite{manakov86c,tang09a}, the nonrelativistic value of $\gamma_0$ is $10913.57(1)$~a.u., and the corresponding relativistic corrections are $-2130(70)$~a.u. (20\%), resulting in $\gamma_0(2p_{3/2})=8784(70)$~a.u. and $-4267(54)$ a.u. (39\%), resulting in $\gamma_0(2p_{1/2})=6647(54)$~a.u. On the other hand, using the present formula, the relativistic corrections on $\gamma_0$ are within the error bars and can be estimated as approximately ~0.01\%. Similarly, the relativistic correction on $\gamma_2$ using the previous formula is 2137(42)~a.u. (29\%), which is significantly larger than the estimated 0.08\% correction obtained from the present formula. Nevertheless, regardless of the magnetic sublevel considered, the overall hyperpolarizability $\gamma$ obtained from both formulas remains consistent. These comparisons indicate that the present formula is more preferable in defining the scalar and tensor components of the hyperpolarizability compared to the previous formula~\cite{manakov86c,tang09a}, especially considering that relativistic effects on physical properties are typically small.

\section{Summary}
In conclusion, we have developed a new formula to calculate the atomic hyperpolarizability, which provides a better separation of scalar and tensor components. This alternative formula has been applied to compute the hyperpolarizabilities for low-lying states of the H, Li, and Be$^+$ systems, both in nonrelativistic and relativistic regimes. Our numerical analysis demonstrates that the relativistic corrections to the scalar and tensor components are not significantly different from their nonrelativistic counterparts, unlike what the previous formula~\cite{manakov86c,tang09a} suggests.

\section{Acknowledgments}
This work was supported by the National Natural Science Foundation of China under Grant Nos. 12174402 and 12393821, by the Strategic Priority Research Program of the Chinese Academy of Sciences under Grant Nos. XDB0920100 and XDB0920101, and by the Nature Science Foundation of Hubei Province Nos. 2019CFA058 and 2022CFA013. ZCY was supported by the Natural Sciences and Engineering Research Council of Canada (NSERC). JFB was supported in part by NSF grant PHY-2116679. All the calculations are finished on the APM-Theoretical Computing Cluster(APM-TCC).

%

%

\appendix

\section{Derivation of the hyperpolarizability}\label {stark_a}

In our previous work, which is referenced as Ref.~\cite{tang09a}, we derived the hyperpolarizability. However, we unintentionally overlooked a crucial constraint involving the Kronecker $\delta(n,0)$. This constraint is necessary when dealing with the second intermediate states $|n\rangle$ with the initial state $|0\rangle$ in the second term of the fourth-order energy correction $\Delta E_4$, as shown in Eqs.~(A34) and (A35) of Ref.~\cite{tang09a}.
In this Appendix, we will revisit the derivation process from the beginning, taking into account the presence of the $\delta(n,0)$ constraint. Remarkably, we have discovered that the new formula we derived is in complete agreement with the previous one, despite having different partitioning in terms of ``$\gamma$" components.

The Hamiltonian of an atom under the influence of a weak external electric field is given by
\begin{equation}
	H = H_0 + H' = H_0 - \bm{\varepsilon} \cdot \bm{d}\,. \label{a1}
\end{equation}
Here, $H_0$ is the atom's unperturbed Hamiltonian, $\bm{\varepsilon}$ is the electric field strength, and $\bm{d}$ is the atom's electric dipole moment, given by $\bm{d} = -e\bm{r}$, where $\bm{r}$ is the electron's position vector relative to a laboratory frame. The perturbation, $H'$, can be expressed as $\bm{r} \cdot \bm{\varepsilon}$ in atomic units. It is noted that the electric quadrupole and magnetic dipole interactions have been omitted from Eq.~(\ref{a1}), and the treatment of these higher-order terms can be found in Refs.~\cite{kolb82,singh18}. 

According to perturbation theory, the fourth-order energy shift~\cite{manakov86c} contains two terms: $\Delta E_4^{(1)}$ and $\Delta E_4^{(2)}$,
\begin{eqnarray}
	\Delta E_4 = \Delta E_4^{(1)} + \Delta E_4^{(2)} = \sum_{mnk}\frac{\langle 0 |H'| m \rangle \langle m |H'| n \rangle \langle n |H'| k \rangle \langle k |H'| 0 \rangle}{(E_0-E_m)(E_0-E_n)(E_0-E_k)} - \sum_{mk}\frac{\langle 0 |H'| m \rangle \langle m |H'| 0 \rangle \langle 0 |H'| k \rangle \langle k |H'| 0 \rangle}{(E_0-E_m)(E_0-E_k)^2}\,. \label{a2}
\end{eqnarray}

Using the spherical tensor operator technique, we can expand $H'| n \rangle \langle n | H'$ of Eq.~(\ref{a2}) as
\begin{eqnarray}
	H'| n \rangle \langle n | H' = \sum_{Kq} (-1)^{K+q} [r^{(1)}\otimes\lambda_nr^{(1)}]_q^K [\varepsilon^{(1)}\otimes\varepsilon^{(1)}]_{-q}^K \,,  \label{a3}
\end{eqnarray}
where $r_\mu^{(1)}=\sqrt{\frac{4\pi}{3}}rY_{1\mu}$ and $\lambda_n=| n \rangle \langle n |$. Assuming that the external electric field $\boldsymbol{\varepsilon}$ is linearly polarized parallel to the $z$ axis, only the $q=0$ component exists in Eq.~(\ref{a3}). Thus we have
\begin{eqnarray}
	H'| n \rangle \langle n | H' &=& \sum\limits_{K} (-1)^{K}\sqrt{2K+1} \sum\limits_{q_1q_2}\begin{pmatrix} 1&1&K\\q_1&q_2&0 \end{pmatrix} r_{q_1}^{(1)}\lambda_nr_{q_2}^{(1)}  [\varepsilon^{(1)}\otimes\varepsilon^{(1)}]_{0}^K\,.    \label{a4}
\end{eqnarray}
Substituting Eq.~(\ref{a4}) into Eq.~(\ref{a2}), $\Delta E_4^{(1)}$ and $\Delta E_4^{(2)}$ can be written as
\begin{eqnarray}
	\Delta E_4^{(1)}&=&\sum\limits_{mnk}\sum\limits_{k_1k_2}\sum\limits_{q_1q_2q_3q_4} (-1)^{k_1+k_2} \sqrt{(2k_1+1)(2k_2+1)} \begin{pmatrix} 1&1&k_1\\ q_1&q_2&0 \end{pmatrix}  \begin{pmatrix} 1&1&k_2\\ q_3&q_4&0 \end{pmatrix} [\varepsilon^{(1)}\otimes\varepsilon^{(1)}]_{0}^{k_1} [\varepsilon^{(1)}\otimes\varepsilon^{(1)}]_{0}^{k_2}    \nonumber \\
	&& \times\frac{\langle 0 |r_{q_1}^{(1)}| m \rangle \langle m |r_{q_2}^{(1)}| n \rangle \langle n
		|r_{q_3}^{(1)}| k \rangle \langle k |r_{q_4}^{(1)}| 0 \rangle}{(E_0-E_m)(E_0-E_n)(E_0-E_k)}\,, \label{a5}
\end{eqnarray}
\begin{eqnarray}
	\Delta E_4^{(2)}&=&-\sum\limits_{m}\sum\limits_{k_1}\sum\limits_{q_1q_2} (-1)^{k_1} \sqrt{2k_1+1} \begin{pmatrix} 1&1&k_1\\ q_1&q_2&0 \end{pmatrix} \frac{\langle 0 |r_{q_1}^{(1)}| m \rangle \langle m |r_{q_2}^{(1)}| 0 \rangle} {(E_0-E_m)} [\varepsilon^{(1)}\otimes\varepsilon^{(1)}]_{0}^{k_1}      \nonumber \\
	&& \times\sum\limits_{k}\sum\limits_{k_2}\sum\limits_{q_3q_4} (-1)^{k_2} \sqrt{2k_2+1} \begin{pmatrix} 1&1&k_2\\ q_3&q_4&0 \end{pmatrix} \frac{\langle 0 |r_{q_3}^{(1)}| k \rangle \langle k |r_{q_4}^{(1)}| 0 \rangle} {(E_0-E_k)^2} [\varepsilon^{(1)}\otimes\varepsilon^{(1)}]_{0}^{k_2}\,,  \label{a6}
\end{eqnarray}
where all the initial and intermediate states are simply denoted as $| 0 \rangle = | n_0JM_J \rangle$, $| m \rangle = | mJ_aM_a \rangle$, $| n \rangle = | nJ_bM_b \rangle$,  $| k \rangle = | kJ_cM_c\rangle$, with $n_0$, $m$, $n$, and $k$ being the corresponding principal quantum numbers of states. The summation in Eqs.~(\ref{a5}) and (\ref{a6}) over $m$, $n$, and $k$ actually represents the summation over all three sets of quantum numbers $\{m,J_a,M_a\}$, $\{n,J_b,M_b\}$, and $\{k,J_c,M_c\}$.

By utilizing the Wigner-Eckart theorem to simplify Eqs.~(\ref{a5}) and (\ref{a6}), we can arrive at the following expressions:
\begin{eqnarray}
	\Delta E_4^{(1)}=-\sum\limits_{J_aJ_bJ_c}\sum\limits_{k_1k_2} (-1)^{k_1+k_2} \sqrt{(2k_1+1)(2k_2+1)} [\varepsilon^{(1)}\otimes\varepsilon^{(1)}]_{0}^{k_1} [\varepsilon^{(1)}\otimes\varepsilon^{(1)}]_{0}^{k_2}  \mathcal{C}_1  \mathcal{T}_1(J_a,J_b,J_c)\,,   \label{a7}
\end{eqnarray}
\begin{eqnarray}
	\Delta E_4^{(2)}=\sum\limits_{J_aJ_c} \sum\limits_{k_1k_2} (-1)^{k_1+k_2} \sqrt{(2k_1+1)(2k_2+1)} [\varepsilon^{(1)}\otimes\varepsilon^{(1)}]_{0}^{k_1} [\varepsilon^{(1)}\otimes\varepsilon^{(1)}]_{0}^{k_2} \mathcal{C}_{2} \mathcal{T}_2(J_a,J_c)\,,   \label{a8}
\end{eqnarray}
where $\mathcal{T}_1(J_a,J_b,J_c)$ and $\mathcal{T}_2(J_a,J_c)$ are radial-dependent terms, $\mathcal{C}_1$ and $\mathcal{C}_2$ are the angular coefficients. The explicit expressions are as follows:
\begin{widetext}
	\begin{eqnarray}
		\mathcal{T}_1(J_a,J_b,J_c)=\sum\limits_{mnk}'\frac{ \langle n_0J \|r^{(1)}\| mJ_a \rangle \langle mJ_a \|r^{(1)}\| nJ_b \rangle \langle nJ_b \|r^{(1)}\| kJ_c \rangle \langle kJ_c \|r^{(1)}\| n_0J \rangle }
		{ \left[E_k(J_c)-E_{n_0}(J)\right] \left[E_m(J_a)-E_{n_0}(J)\right]  \left[E_n(J_b)-E_{n_0}(J)\right] }\,,   \label{a9}
	\end{eqnarray}
	\begin{eqnarray}
		\mathcal{T}_2(J_a,J_c)=\sum\limits_{m}'\frac{ \langle n_0J \|r^{(1)}\| mJ_a \rangle \langle mJ_a \|r^{(1)}\| n_0J \rangle }{\left[E_m(J_a)-E_{n_0}(J)\right]}
		\sum\limits_{k}'\frac{ \langle n_0J \|r^{(1)}\| kJ_c \rangle \langle kJ_c \|r^{(1)}\| n_0J \rangle }{\left[E_k(J_c)-E_{n_0}(J)\right]^2}\,,  \label{a10}
	\end{eqnarray}
	\begin{eqnarray}
		\mathcal{C}_1&=&\sum\limits_{M_aM_bM_c}\sum\limits_{q_1q_2q_3q_4} (-1)^{J-M_J+J_a-M_a+J_b-M_b+J_c-M_c}  \begin{pmatrix} 1&1&k_1\\ q_1&q_2&0 \end{pmatrix}  \begin{pmatrix} 1&1&k_2\\ q_3&q_4&0 \end{pmatrix} \begin{pmatrix} J&1&J_a\\-M_J&q_1&M_a \end{pmatrix}  \nonumber   \\
		& &\times \begin{pmatrix} J_a&1&J_b\\-M_a&q_2&M_b \end{pmatrix} \begin{pmatrix} J_b&1&J_c\\-M_b&q_3&M_c \end{pmatrix}   \begin{pmatrix} J_c&1&J\\-M_c&q_4&M_J \end{pmatrix}\,,  \label{a11}
	\end{eqnarray}
	\begin{eqnarray}
		\mathcal{C}_{2}&=&\sum\limits_{M_aM_c}\sum\limits_{q_1q_2q_3q_4}  (-1)^{J_a-M_a+J_c-M_c}  \begin{pmatrix} 1&1&k_1\\ q_1&q_2&0 \end{pmatrix}  \begin{pmatrix} J&1&J_a\\-M_J&q_1&M_a \end{pmatrix}  \begin{pmatrix} J_a&1&J\\-M_a&q_2&M_J \end{pmatrix} \begin{pmatrix} 1&1&k_2\\ q_3&q_4&0 \end{pmatrix}  \nonumber   \\
		& &\times \begin{pmatrix} J&1&J_c\\-M_J&q_3&M_c \end{pmatrix}  \begin{pmatrix} J_c&1&J\\-M_c&q_4&M_J \end{pmatrix}\,, \label{a12}
	\end{eqnarray}
\end{widetext}
where the primes over the summations in Eqs.~(\ref{a9}) and (\ref{a10}) mean the omission of any intermediate states that are degenerate with the initial state, and the summations over intermediate states include the negative-energy states in the relativistic case.
By using the graphical method of angular momentum, we can simplify $\mathcal{C}_{1}$ and $\mathcal{C}_{2}$:
\begin{eqnarray}
	\mathcal{C}_1&=&(-1)^{J-M_J} \begin{Bmatrix} 1&1&k_1\\J&J_b&J_a \end{Bmatrix}  \begin{Bmatrix} 1&1&k_2\\J&J_b&J_c \end{Bmatrix} \sum\limits_{\Lambda}(2\Lambda+1)
	\begin{pmatrix} J&J&\Lambda\\-M_J&M_J&0 \end{pmatrix}  \begin{pmatrix} k_1&k_2&\Lambda\\0&0&0 \end{pmatrix}  \begin{Bmatrix} k_1&k_2&\Lambda\\J&J&J_b \end{Bmatrix}\,,  \label{a13}
\end{eqnarray}
\begin{eqnarray}
	\mathcal{C}_2&=& \begin{pmatrix} J&J&k_1\\-M_J&M_J&0 \end{pmatrix} \begin{pmatrix} J&J&k_2\\-M_J&M_J&0 \end{pmatrix} \begin{Bmatrix} 1&1&k_1\\J&J&J_a \end{Bmatrix} \begin{Bmatrix} 1&1&k_2\\J&J&J_c \end{Bmatrix}\,.   \label{a14}
\end{eqnarray}
Since $\boldsymbol{\varepsilon}$ is oriented along the $z$ axis, we can express $[\varepsilon^{(1)}\otimes\varepsilon^{(1)}]_{0}^{k}$ in the form:
\begin{eqnarray}
	[\varepsilon^{(1)}\otimes\varepsilon^{(1)}]_{0}^{k}=\sqrt{2k+1} \begin{pmatrix} 1&1&k\\0&0&0 \end{pmatrix} \varepsilon^2\,. \label{a15}
\end{eqnarray}
By substituting Eqs.~(\ref{a13}), (\ref{a14}), and (\ref{a15}) into Eqs.~(\ref{a7}) and (\ref{a8}), we obtain the final expressions for $\Delta E_4^{(1)}$ and $\Delta E_4^{(2)}$:
\begin{eqnarray}
	\Delta E_4^{(1)}&=&-\varepsilon^4 \sum\limits_{J_aJ_bJ_c} \mathcal{T}_1(J_a,J_b,J_c) \sum\limits_{\lambda}(-1)^{J-M_J} \begin{pmatrix} J&J&\lambda\\-M_J&M_J&0 \end{pmatrix} \mathcal{G}_\lambda^{(1)}(J,J_a,J_b,J_c)\,,  \label{a16}  \\ \notag
\end{eqnarray}
\begin{eqnarray}
	\Delta E_4^{(2)}&=&\varepsilon^4 \sum\limits_{J_aJ_c} \mathcal{T}_2(J_a,J_c) \sum\limits_{k_1k_2} \begin{pmatrix} J&J&k_1\\-M_J&M_J&0 \end{pmatrix}  \begin{pmatrix} J&J&k_2\\-M_J&M_J&0 \end{pmatrix} \mathcal{G}_{k_1k_2}^{(2)}(J,J_a,J_c)\,,  \label{a17}
\end{eqnarray}
where
\begin{eqnarray}
	\mathcal{G}_\lambda^{(1)}(J,J_a,J_b,J_c)&=&\sum\limits_{k_1k_2} (\lambda,k_1,k_2)  \begin{pmatrix} 1&1&k_1\\0&0&0 \end{pmatrix}  \begin{pmatrix} 1&1&k_2\\0&0&0 \end{pmatrix}   \begin{pmatrix} k_1&k_2&\lambda\\0&0&0 \end{pmatrix} \begin{Bmatrix} 1&1&k_1\\J&J_b&J_a \end{Bmatrix} \begin{Bmatrix} 1&1&k_2\\J&J_b&J_c \end{Bmatrix} \begin{Bmatrix} k_1&k_2&\lambda\\J&J&J_b \end{Bmatrix}\,,   \nonumber \\
	\label{a18}
\end{eqnarray}
\begin{eqnarray}
	\mathcal{G}_{k_1k_2}^{(2)}(J,J_a,J_c)&=&(k_1,k_2) \begin{pmatrix} 1&1&k_1\\0&0&0 \end{pmatrix}  \begin{pmatrix} 1&1&k_2\\0&0&0 \end{pmatrix}  \begin{Bmatrix} 1&1&k_1\\J&J&J_a \end{Bmatrix} \begin{Bmatrix} 1&1&k_2\\J&J&J_c \end{Bmatrix}\,,  \label{a19}
\end{eqnarray}
with the notation $(\lambda,k_1,k_2)=(2\lambda+1)(2k_1+1)(2k_2+1)$, and similarly for $(k_1,k_2)$.

According to the properties of the $3j$ symbol, the permissible values for $k_1$ and $k_2$ in Eqs.~(\ref{a18}) and (\ref{a19}) are 0 and 2, while the permissible values for $\lambda$ are 0, 2, and 4.
Also using the following formulas:
\begin{eqnarray}
	(-1)^{J-M_J}\begin{pmatrix} J&J&0\\-M_J&M_J&0 \end{pmatrix} = (-1)^{2J}\frac{1}{\sqrt{2J+1}}\,, \label{a20}
\end{eqnarray}
\begin{eqnarray}
	(-1)^{J-M_J}\begin{pmatrix} J&J&2\\-M_J&M_J&0 \end{pmatrix} = (-1)^{2J}\frac{3M_J^2-J(J+1)}{\sqrt{(2J+3)(J+1)(2J+1)J(2J-1)}}\,,  \label{a21}
\end{eqnarray}
\begin{eqnarray}
	(-1)^{J-M_J}\begin{pmatrix} J&J&4\\-M_J&M_J&0 \end{pmatrix} = (-1)^{2J}\frac{3(5M_J^2-J^2-2J)(5M_J^2+1-J^2)-10M_J^2(4M_J^2-1)}{\sqrt{(2J+5)(J+2)(2J+3)(J+1)(2J+1)2J(2J-1)(2J-2)(2J-3)}}\,,  \label{a22}
\end{eqnarray}
the fourth-order energy shift $\Delta E_4$ can finally be simplified in the form:
\begin{eqnarray}
	\Delta E_4=\Delta E_4^{(1)}+\Delta E_4^{(2)}=-\frac{\varepsilon^4}{24}[\gamma^{(1)}+\gamma^{(2)}]\,, \label{a23}
\end{eqnarray}
where
\begin{eqnarray}
	\gamma^{(1)}=\gamma_0^{(1)}+g_2(J,M_J)\gamma_2^{(1)}+g_4(J,M_J)\gamma_4^{(1)}\,,  \label{a24}
\end{eqnarray}
\begin{eqnarray}
	\gamma^{(2)}=\gamma_0^{(2)}+g_2(J,M_J)\gamma_2^{(2)}+g_2^2(J,M_J)\gamma_4^{(2)}\,. \label{a25}
\end{eqnarray}
By denoting $\gamma$ as the total hyperpolarizability, we obtain
\begin{widetext}
	\begin{eqnarray}
		\gamma&=&\gamma^{(1)}+\gamma^{(2)}=\gamma_0+g_2(J,M_J)\gamma_2+[g_4(J,M_J)\gamma_4^{(1)}+g_2^2(J,M_J)\gamma_4^{(2)}]\,,  \label{a26}
	\end{eqnarray}
\end{widetext}
where $\gamma_0=\gamma_0^{(1)}+\gamma_0^{(2)}$, $\gamma_2=\gamma_2^{(1)}+\gamma_2^{(2)}$, and the coefficients $g_2(J,M_J)$ and $g_4(J,M_J)$ depend also on the value of $M_J$:
\begin{eqnarray}
	g_2(J,M_J)=
	\begin{cases}
		0,                            & J \leq \frac{1}{2}\,, \\
		\frac{3M_J^2-J(J+1)}{J(2J-1)},& {\rm otherwise}\,,      \label{a27}
	\end{cases}
\end{eqnarray}
\begin{eqnarray}
	g_4(J,M_J)=
	\begin{cases}
		0,                            & J \leq \frac{3}{2}\,, \\
		\frac{3(5M_J^2-J^2-2J)(5M_J^2+1-J^2)-10M_J^2(4M_J^2-1)}{J(2J-1)(2J-2)(2J-3)},& {\rm otherwise}\,.      \label{a28}
	\end{cases}
\end{eqnarray}
The values of $\gamma_0^{(1)}$ and $\gamma_0^{(2)}$ correspond to the scalar components of the second hyperpolarizability, whereas $\gamma_2^{(1)}$, $\gamma_2^{(2)}$, $\gamma_4^{(1)}$, and $\gamma_4^{(2)}$ correspond to the tensor components, which have the following forms:
\begin{eqnarray}
	\gamma_0^{(1)}&=&(-1)^{2J}\frac{24}{\sqrt{2J+1}}  \sum\limits_{J_aJ_bJ_c}\mathcal{G}_0^{(1)}(J,J_a,J_b,J_c)\mathcal{T}_1(J_a,J_b,J_c)\,,  \label{a29}
\end{eqnarray}
%
\begin{eqnarray}
	\gamma_2^{(1)}&=&(-1)^{2J} 24\sqrt{\frac{J(2J-1)}{(2J+3)(J+1)(2J+1)}} \sum\limits_{J_aJ_bJ_c}\mathcal{G}_2^{(1)}(J,J_a,J_b,J_c)\mathcal{T}_1(J_a,J_b,J_c)\,,  \label{a30}
\end{eqnarray}
%
\begin{eqnarray}
	\gamma_4^{(1)}&=&(-1)^{2J} 24\sqrt{\frac{J(2J-1)(J-1)(2J-3)}{(2J+5)(J+2)(2J+3)(J+1)(2J+1)}}  \sum\limits_{J_aJ_bJ_c}\mathcal{G}_4^{(1)}(J,J_a,J_b,J_c)\mathcal{T}_1(J_a,J_b,J_c)\,.   \label{a31}
\end{eqnarray}
%
%
\begin{eqnarray}
	\gamma_0^{(2)}=-\frac{24}{2J+1}\sum\limits_{J_aJ_c}\mathcal{G}_{00}^{(2)}(J,J_a,J_c) \mathcal{T}_2(J_a,J_c)\,,   \label{a32}
\end{eqnarray}
%
\begin{eqnarray}
	\gamma_2^{(2)}&=&-\frac{24}{2J+1} \sqrt{\frac{J(2J-1)}{(2J+3)(J+1)}}     \sum\limits_{J_aJ_c} [\mathcal{G}_{02}^{(2)}(J,J_a,J_c)+\mathcal{G}_{20}^{(2)}(J,J_a,J_c)]  \mathcal{T}_2(J_a,J_c)\,, \label{a33}
\end{eqnarray}
%
\begin{eqnarray}
	\gamma_4^{(2)}&=&-\frac{24J(2J-1)}{(2J+3)(2J+1)(J+1)} \sum\limits_{J_aJ_c}\mathcal{G}_{22}^{(2)}(J,J_a,J_c)  \mathcal{T}_2(J_a,J_c)\,.  \label{a34}
\end{eqnarray}
%
In the nonrelativistic case, we only need to substitute the total angular quantum number $J$ with the orbital angular quantum number $L$, and replace the magnetic quantum number $M_J$ with $M_L$.

\section{Convergence test for excited states of the H atom}\label {table_a}

\setcounter{table}{0}
\renewcommand\thetable{S\arabic{table}}	

\begingroup
\squeezetable
\begin{table*}[!htbp]
\caption{\label{ta1} Convergence of the nonrelativistic pure state hyperpolarizabilities $\gamma_0$, $\gamma_2$, and $\gamma_4^{(2)}$ for the $2p$ state of the H atom, in atomic units.}
\begin{ruledtabular}
\begin{tabular}{llll}
\multicolumn{1}{l}{$N$}&\multicolumn{1}{c}{$10^{-6}\gamma_0$}&\multicolumn{1}{c}{$10^{-6}\gamma_2$}&\multicolumn{1}{c}{$10^{-4}\gamma_4^{(2)}$}\\ \hline
100&  8.130 559 999 999 999 42&           $-$2.769 471 999 999 997 9&           $-$3.423 999 999 999 996 7\\
150&  8.130 559 999 999 999 996 5&        $-$2.769 471 999 999 999 987&         $-$3.423 999 999 999 999 981\\
200&  8.130 559 999 999 999 999 899&      $-$2.769 471 999 999 999 999 64&      $-$3.423 999 999 999 999 999 47\\
250&  8.130 559 999 999 999 999 993 5&    $-$2.769 471 999 999 999 999 977&     $-$3.423 999 999 999 999 999 966\\
300&  8.130 559 999 999 999 999 999 30&   $-$2.769 471 999 999 999 999 997 5&   $-$3.423 999 999 999 999 999 996 4\\
350&  8.130 559 999 999 999 999 999 89&   $-$2.769 471 999 999 999 999 999 62&  $-$3.423 999 999 999 999 999 999 45\\
400&  8.130 559 999 999 999 999 999 98&   $-$2.769 471 999 999 999 999 999 93&  $-$3.423 999 999 999 999 999 999 89\\
Extrap.& 8.130 560  &$-$2.769 472 & $-$3.424 0\\
\end{tabular}
\end{ruledtabular}
\end{table*}
\endgroup
\begingroup
\squeezetable
\begin{table*}[!htbp] 
\caption{\label{ta2} Convergence of the nonrelativistic pure state hyperpolarizabilities $\gamma_0$, $\gamma_2$, $\gamma_4^{(1)}$, and $\gamma_4^{(2)}$ for the $3d$ state of the H atom, in atomic units.}
\begin{ruledtabular}
\begin{tabular}{lllll}
\multicolumn{1}{l}{$N$}&\multicolumn{1}{c}{$10^{-9}\gamma_0$}&\multicolumn{1}{c}{$10^{-9}\gamma_2$}&\multicolumn{1}{c}{$10^{-7}\gamma_4^{(1)}$}&\multicolumn{1}{c}{$10^{-7}\gamma_4^{(2)}$}\\  \hline
100&  1.913 524 179 299 999 1&           $-$1.027 110 959 357 141 8&           2.939 248 330 714 278&          $-$6.944 162 399 999 999 0\\
150&  1.913 524 179 299 999 995&         $-$1.027 110 959 357 142 851&         2.939 248 330 714 285 67&       $-$6.944 162 399 999 999 994\\
200&  1.913 524 179 299 999 999 85&      $-$1.027 110 959 357 142 857 0&       2.939 248 330 714 285 713&      $-$6.944 162 399 999 999 999 85\\
250&  1.913 524 179 299 999 999 991&     $-$1.027 110 959 357 142 857 13&      2.939 248 330 714 285 714 20&   $-$6.944 162 399 999 999 999 990\\
300&  1.913 524 179 299 999 999 999 0&   $-$1.027 110 959 357 142 857 141 6&   2.939 248 330 714 285 714 277&  $-$6.944 162 399 999 999 999 999 0\\
350&  1.913 524 179 299 999 999 999 85&  $-$1.027 110 959 357 142 857 142 7&   2.939 248 330 714 285 714 284&  $-$6.944 162 399 999 999 999 999 85\\
400&  1.913 524 179 299 999 999 999 97&  $-$1.027 110 959 357 142 857 142 8&   2.939 248 330 714 285 714 285&  $-$6.944 162 399 999 999 999 999 97\\		
Extrap& 1.913 524 179 3               &  $-$1.027 110 959 357 142 857 142 8&   2.939 248 330 714 285 714 286&  $-$6.944 162 4 \\
\end{tabular}
\end{ruledtabular}
\end{table*}
\endgroup
\begingroup
\squeezetable
\begin{table*}[!htbp]
\caption{\label{ta3} Convergence of the relativistic pure state hyperpolarizabilities $\gamma_0$, $\gamma_2$, and $\gamma_4^{(2)}$ for the $2p_{1/2}$ and $2p_{3/2}$ states of the H atom, in atomic units.}
\begin{ruledtabular}
\begin{tabular}{lllll}
\multirow{2}*{$N$}&\multicolumn{1}{c}{$2p_{1/2}$}&\multicolumn{3}{c}{$2p_{3/2}$}\\
\cline{2-2} \cline{3-5}	
&\multicolumn{1}{c}{$10^{-6}\gamma_0$}&\multicolumn{1}{c}{$10^{-6}\gamma_0$}&\multicolumn{1}{c}{$10^{-6}\gamma_2$}&\multicolumn{1}{c}{$10^{-4}\gamma_4^{(2)}$}\\  \hline																						
100&     8.129 480 180 856            &  8.130 501 053 31             &  $-$2.769 565 818 758            &  $-$3.424 116 618 84            \\           
200&     8.129 480 180 847 221        &  8.130 501 053 270 538 9      &  $-$2.769 565 818 725 861        &  $-$3.424 116 618 636 726 0     \\       
300&     8.129 480 180 847 218 663    &  8.130 501 053 270 535 691    &  $-$2.769 565 818 725 858 675    &  $-$3.424 116 618 636 721 347   \\   
400&     8.129 480 180 847 218 657 977&  8.130 501 053 270 535 680 91 &  $-$2.769 565 818 725 858 667 51 &  $-$3.424 116 618 636 721 334 86 \\
500&     8.129 480 180 847 218 657 893&  8.130 501 053 270 535 680 760&  $-$2.769 565 818 725 858 667 395&  $-$3.424 116 618 636 721 334 673\\
600&     8.129 480 180 847 218 657 892&  8.130 501 053 270 535 680 756&  $-$2.769 565 818 725 858 667 393&  $-$3.424 116 618 636 721 334 669\\
Extrap.& 8.129 480 180 847 218 657 892&  8.130 501 053 270 535 680 756&  $-$2.769 565 818 725 858 667 393&  $-$3.424 116 618 636 721 334 669\\
\end{tabular}
\end{ruledtabular}
\end{table*}
\endgroup
\begingroup
\squeezetable
\begin{table*}[!htbp]
\caption{\label{ta4} Convergence of the relativistic hyperpolarizabilities pure state $\gamma_0$, $\gamma_2$, and $\gamma_4^{(2)}$ for the $3d_{3/2}$ state of the H atom, in atomic units.}
\begin{ruledtabular}
\begin{tabular}{llll}
\multicolumn{1}{l}{$N$}&\multicolumn{1}{c}{$10^{-9}\gamma_0$}&\multicolumn{1}{c}{$10^{-8}\gamma_2$}&\multicolumn{1}{c}{$10^{-7}\gamma_4^{(2)}$}\\  \hline
100&      1.913 481 814 79            &  $-$7.189 813 884               & $-$3.402 556 954 7             \\
200&      1.913 481 814 288 678       &  $-$7.189 813 879 847 98        & $-$3.402 556 953 695 364       \\
300&      1.913 481 814 288 667 955   &  $-$7.189 813 879 847 891 83    & $-$3.402 556 953 695 351 874   \\
400&      1.913 481 814 288 667 929 1 &  $-$7.189 813 879 847 891 586 0 & $-$3.402 556 953 695 351 845 92\\
500&      1.913 481 814 288 667 928 69&  $-$7.189 813 879 847 891 582 22& $-$3.402 556 953 695 351 845 49\\
600&      1.913 481 814 288 667 928 68&  $-$7.189 813 879 847 891 582 07& $-$3.402 556 953 695 351 845 47\\
Extrap.&  1.913 481 814 288 667 928 67&  $-$7.189 813 879 847 891 582 06& $-$3.402 556 953 695 351 845 47\\
\end{tabular}
\end{ruledtabular}
\end{table*}
\endgroup
\begingroup
\squeezetable
\begin{table*}[!htbp]
\caption{\label{ta5} Convergence of the relativistic pure state hyperpolarizabilities $\gamma_0$, $\gamma_2$, $\gamma_4^{(1)}$, and $\gamma_4^{(2)}$ for the $3d_{5/2}$ state of the H atom, in atomic units.}
\begin{ruledtabular}
\begin{tabular}{lllll}
\multicolumn{1}{l}{$N$}&\multicolumn{1}{c}{$10^{-9}\gamma_0$} &\multicolumn{1}{c}{$10^{-9}\gamma_2$}&\multicolumn{1}{c}{$10^{-7}\gamma_4^{(1)}$}&\multicolumn{1}{c}{$10^{-7}\gamma_4^{(2)}$}\\  \hline
100&     1.913 513 569 27            &  $-$1.027 113 325 40            &  2.939 326 813 538           &  $-$6.944 098 826 2             \\
200&     1.913 513 569 010 927 2     &  $-$1.027 113 325 161 428 2     &  2.939 326 812 562 20        &  $-$6.944 098 825 526 127       \\
300&     1.913 513 569 010 920 534   &  $-$1.027 113 325 161 421 963   &  2.939 326 812 562 169 80    &  $-$6.944 098 825 526 114 456   \\
400&     1.913 513 569 010 920 516 1 &  $-$1.027 113 325 161 421 946 33&  2.939 326 812 562 169 713 4 &  $-$6.944 098 825 526 114 424 88\\
500&     1.913 513 569 010 920 515 82&  $-$1.027 113 325 161 421 946 06&  2.939 326 812 562 169 711 98&  $-$6.944 098 825 526 114 424 39\\
600&     1.913 513 569 010 920 515 81&  $-$1.027 113 325 161 421 946 05&  2.939 326 812 562 169 711 92&  $-$6.944 098 825 526 114 424 37\\
Extrap.& 1.913 513 569 010 920 515 81&  $-$1.027 113 325 161 421 946 05&  2.939 326 812 562 169 711 92&  $-$6.944 098 825 526 114 424 37\\
\end{tabular}
\end{ruledtabular}
\end{table*}
\endgroup

\end{document}